\documentclass[11pt]{article}
\usepackage[margin=0.88in]{geometry}
\usepackage[T1]{fontenc}
\usepackage{lmodern,amsmath,amssymb,bm,graphicx,booktabs,microtype,hyperref,fancyhdr}
\usepackage{authblk}
\hypersetup{colorlinks=true,linkcolor=blue,citecolor=blue,urlcolor=blue}
\hypersetup{hidelinks}
\newcommand{\B}{\mathbf B}\newcommand{\bb}{\mathbf b}\newcommand{\xx}{\boldsymbol\xi}
\newcommand{\PP}{\mathsf P}\newcommand{\I}{\mathsf I}
\begin{document}

\pagestyle{fancy}\fancyhf{}\lhead{Exact anisotropic confinement}\rhead{}\cfoot{\thepage}
\setlength{\headheight}{14pt}
\title{Exact analytic anisotropic three-dimensional equilibria with nested magnetic surfaces, confined drift orbits and without spatial symmetry}
\date{}
\author[1]{A. Bhattacharjee}
\author[1]{W. Sengupta}
\author[2,1]{V. N. Duarte}
\author[3]{L. Fu}
\affil[1]{Department of Astrophysical Sciences, Princeton University, Princeton, New Jersey 08543, USA}
\affil[2]{Princeton Plasma Physics Laboratory, Princeton University, Princeton, New Jersey 08543, USA}
\affil[3]{Courant Institute of Mathematical Sciences, New York University, New York, New York 10012, USA}

\maketitle

\begin{abstract}
Exact toroidal fluid and kinetic equilibria with nested magnetic surfaces and no continuous symmetry are of great interest in laboratory, astrophysical and space plasmas. Their existence and stability have been open in physics and mathematics for decades. We construct such anisotropic-pressure equilibria by applying the Clemente transformation to exact vacuum fields. They connect smoothly to these fields on a fixed regular domain and preserve a regular magnetic axis, perfectly nested surfaces and spatially varying rotational transform. To our knowledge, the sheared configurations are the first exact anisotropic equilibria combining these properties with positive pressures and a positive firehose margin. The fixed-boundary Chew--Goldberger--Low (CGL) quadratic energy splits into a nonnegative vacuum-field term and a local pressure term, nonnegative for $p_\perp/p_\parallel\le(3+\sqrt{21})/2\approx3.79$. For $p_\parallel\ge p_\perp$ each configuration minimizes the finite-amplitude CGL energy over smooth boundary-fixed ideal material deformations. For the current-free configurations, physical guiding-center distributions produce these pressures within the same bound, with no free energy for magnetic-moment-conserving electrostatic perturbations. Trapped particles inherit the vacuum field's second adiabatic invariant at a rescaled magnetic moment, so local closed action contours carry over. The equilibrium code DESC reproduces them from the exact solution. Kinetic stability for surface-dependent rescalings and against general nonlinear electromagnetic perturbations remains open.
\end{abstract}

Magnetic confinement fusion aims to hold a plasma hotter than the core of the
Sun inside a toroidal magnetic field. The field lines wind around nested
toroidal surfaces, and charged particles stream rapidly along them. Two
concepts lead the field. The tokamak is axisymmetric and drives a large
current through the plasma to twist its field lines. The stellarator twists
them with external coils alone, at the cost of a fully three-dimensional (3D)
geometry~\cite{Helander2014}. Symmetry has guided the design of both. In a
tokamak, axisymmetry gives every particle a conserved canonical angular
momentum, which keeps its slow drifts close to a magnetic surface on which field lines lie. A generic
stellarator lacks this protection, and its trapped particles drift out much
faster. The way forward was quasisymmetry, which is a type of hidden symmetry: if the field strength, though not the field's shape, is symmetric in suitable coordinates, guiding centres are
confined as well as in a tokamak~\cite{Boozer1983,NuhrenbergZille1988,LandremanPaul2022}.
A weaker principle, omnigeneity, asks only that the bounce-averaged drift
across the surfaces vanish~\cite{HallMcnamara1975,CaryShasharina1997}, and
reduced neoclassical transport has now been demonstrated in an optimized
stellarator~\cite{Beidler2021}. Here we ask whether a spatial symmetry is
needed at all. We study fields without continuous symmetry whose trapped particles nevertheless drift on local closed contours of
the second adiabatic invariant. Such
fields trade an exact symmetry of the field for a property of the particle
orbits themselves.

The existence of smooth 3D equilibria with nested flux surfaces has remained a long-standing problem in plasma as well as mathematical physics. Grad argued that, with scalar pressure, such equilibria generally require a continuous symmetry~\cite{grad}. The issue is more pronounced for quasisymmetric fields, which confine guiding centers as well as an axisymmetric field does. Garren and Boozer showed that the expansion of a quasisymmetric equilibrium about its magnetic axis is overdetermined ~\cite{garren,landremansengupta}. However, overdetermined equations can have solutions. Exact equilibria that sharpen these questions have recently been constructed. They include 3D equilibria with scalar pressure and nested surfaces~\cite{gomezserrano,landreman}, some without stellarator symmetry~\cite{issan} or with an arbitrary number of field periods~\cite{issannperiod}, and vacuum fields with nested tori and no continuous symmetry~\cite{sengupta}.

Real stellarator plasmas need not have scalar pressure. Neutral-beam injection and ion-cyclotron heating create fast-ion populations with very different parallel and perpendicular pressures, and the guiding-center description of a collisionless plasma gives a gyrotropic pressure tensor ~\cite{cgl}. 3D equilibrium codes with anisotropic pressure have been developed and deployed ~\cite{cooper, dudt, panici}. Anisotropy also bears directly on the equilibrium existence question. Rodr\'iguez and Bhattacharjee ~\cite{rodriguez} imposed quasisymmetry on the Kruskal--Kulsrud variational principle~\cite{kruskal} through a Lagrange multiplier and found that the multiplier acts as an anisotropic pressure. A recent variational treatment of guiding-center kinetics~\cite{fubhattacharjee} showed that the pressures of any kinetic equilibrium built on the invariants of guiding-center motion satisfy parallel force balance identically at the kinetic as well as the fluid level of description, without imposing parallel force balance through an additional Lagrange multiplier.

The existence of an exact finite-pressure equilibrium leaves a separate question: can it be connected continuously to a nonzero vacuum field on a fixed regular toroidal domain? The scalar-pressure families in Ref.~\cite{landreman} demonstrate exact nested-surface equilibria and axisymmetric limits. That work does not establish a continuation to a nonzero vacuum field on a fixed regular toroidal domain.

\begin{figure}[tbp]\centering\includegraphics[width=.94\linewidth]{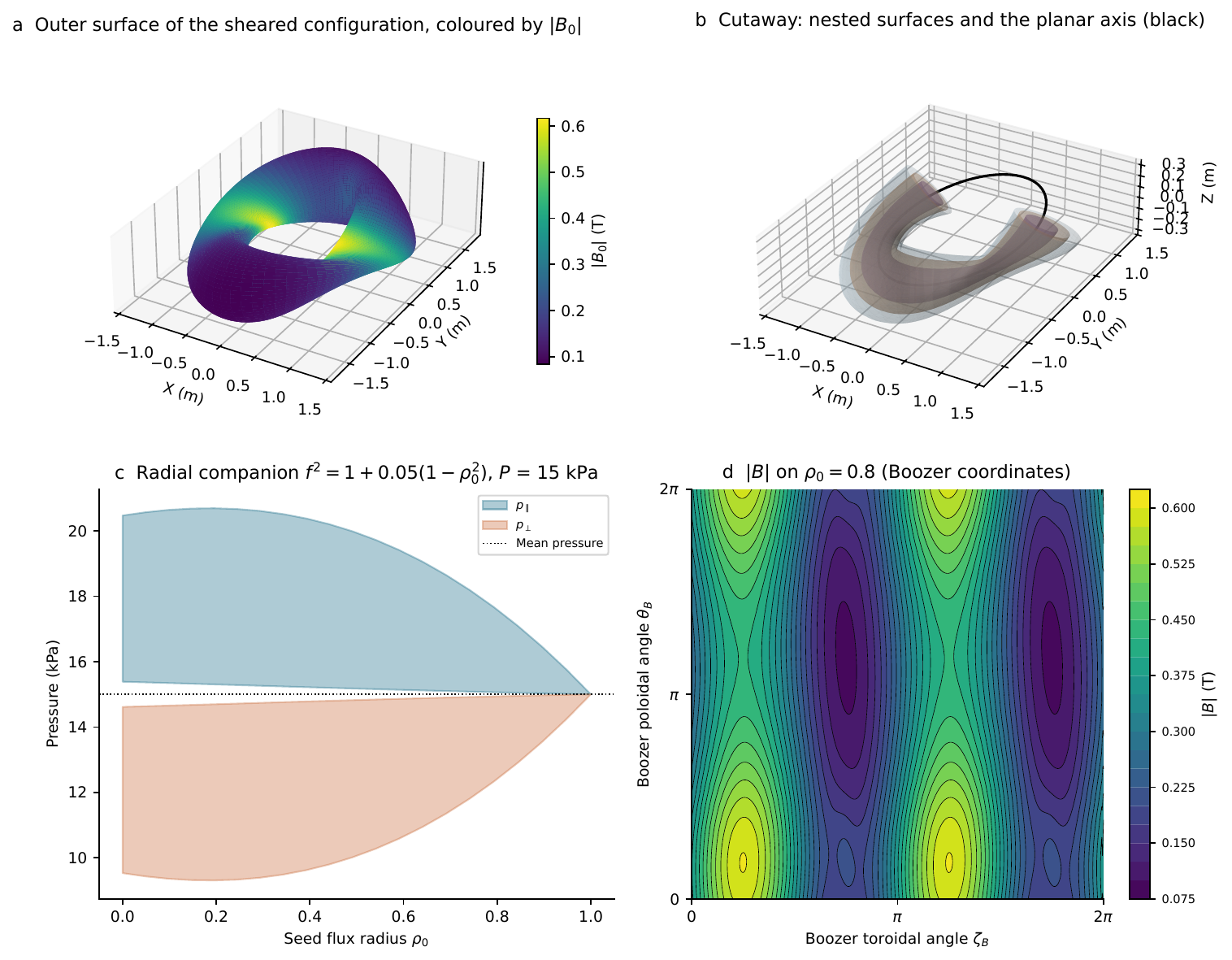}
\caption{\textbf{A nonsymmetric ellipsoidal configuration with finite magnetic shear.} \textbf{a}, Outer surface of the $N=1$ ellipsoidal vacuum seed, coloured by the exact field strength. \textbf{b}, Cutaway showing nested surfaces and the planar noncircular magnetic axis (black). \textbf{c}, Pressure envelopes of the current-carrying companion $f^2=1+0.05(1-\rho_0^2)$ with $P=15$ kPa, sampled over each surface. \textbf{d}, Contours of the seed field strength $|B|$ in Boozer coordinates at $\rho_0=0.8$. The companions keep these surfaces and the rotational transform, which falls by $7.9\%$ from axis to edge (Fig.~\ref{sfig:sheared}). This configuration is distinct from the $N=3$ action example of Fig.~\ref{fig:action}.}\label{fig:sheared}
\end{figure}

In this paper, we construct exact nonsymmetric toroidal anisotropic equilibria by applying a classical pressure-field transformation to the new vacuum seeds of Ref.~\cite{sengupta}. They connect continuously to 3D nonsymmetric vacuum fields, have perfectly nested flux surfaces through the whole volume with a regular magnetic axis, sheared rotational transform, positive pressures, and no continuous symmetry. A single parameter switches on the anisotropy, the plasma current and the Lorentz force, while the flux surfaces and the transform remain exactly those of the vacuum field. Shear depends on the seed. We show an ellipsoidal configuration with finite magnetic shear (Fig.~\ref{fig:sheared} and Appendix~\ref{app:1}) and a second, also sheared, ellipsoidal configuration whose trapped particles have local closed action contours. A shearless paraboloidal benchmark is treated in Methods and the appendices. To our knowledge, the sheared ellipsoidal companions are the first exact anisotropic equilibria that combine nested toroidal magnetic surfaces about a regular axis, given in closed form, with nonzero rotational transform, finite magnetic shear, no continuous symmetry, positive pressures, a positive firehose margin and a continuous connection to a nonzero vacuum field. None of the earlier constructions cited below establish all these properties together, which bring these equilibria a step closer to physical realizability. Three consequences follow, for ideal stability, for kinetic pressure support and for drift confinement.

\subsection*{Adding pressure while preserving exact surfaces}
The static equilibrium of a plasma with gyrotropic pressure tensor $\PP=p_\perp\I+(p_\parallel-p_\perp)\bb\bb$ satisfies~\cite{cgl}
\begin{equation}
\nabla\cdot\PP=\mathbf j\times\B,\qquad \mu_0\mathbf j=\nabla\times\B,\qquad \nabla\cdot\B=0,
\label{static}
\end{equation}
equivalently $\nabla\cdot\bigl(\PP+B^2\I/(2\mu_0)-\B\B/\mu_0\bigr)=0$ with $\nabla\cdot\B=0$.

Exact solutions of equations~\eqref{static} follow from any smooth isotropic magnetostatic equilibrium $(\B_0,p_0)$ with $\B_0\ne0$, $\mu_0\mathbf j_0=\nabla\times\B_0$ and $\nabla p_0=\mathbf j_0\times\B_0$. We take a function $f>0$ with $\B_0\cdot\nabla f=0$ and set $Q=(f^2-1)B_0^2/(2\mu_0)$. For any constant $C$, the fields $\B=f\B_0$, $p_\parallel=p_0+C+Q$ and $p_\perp=p_0+C-Q$ solve equations~\eqref{static} exactly, because the total stress equals the seed's isotropic stress plus $C\:\I$. The current is $\mathbf j=f\mathbf j_0+\nabla f\times\B_0/\mu_0$, and both pressures are positive where $p_0+C>|Q|$. This is the classical transformation introduced by Clemente for axisymmetric equilibria~\cite{clemente} and given in three dimensions by Cheviakov, Bogoyavlenskij and Anco~\cite{cheviakov,anco}. Because $f$ is constant on field lines, the map keeps the seed's field lines, magnetic surfaces and rotational transform.

Applied to the exact 3D scalar-pressure equilibria with nested toroidal surfaces of Refs.~\cite{gomezserrano,landreman,issan,issannperiod}, the map gives exact nonsymmetric anisotropic equilibria with nested surfaces and the seeds' rotational transform. No continuation of those seeds to a nonzero vacuum field on a fixed domain has been established. Some earlier constructions have sought solutions of the kind given here in different ways, and we state how each differs. Cheviakov applied the map to seeds whose magnetic surfaces are helical cylinders, ellipsoidal shells or open flux tubes~\cite{cheviakov,cheviakovsurf}. His thesis also applied it to the vacuum field of a nonsymmetric closed current loop outside the plasma, whose field lines were found by numerical tracing to cover nested tori~\cite{cheviakovthesis}. That is the closest earlier construction, but its surfaces and rotational transform were not established analytically. Sato and Yamada~\cite{satoyamada} rescaled the harmonic field $\nabla(\varphi+\epsilon x)$ by a function of an explicit family of nested tori with no continuous Euclidean symmetry, which is the same map with a vacuum seed. That seed has no vertical component, so every field line is a closed planar curve and the rotational transform is zero. Its tori are degenerate and only a choice among many invariant families rather than surfaces fixed by the field. Exact anisotropic constructions with exactly zero firehose margin, and related flowing equilibria, are compared in Methods.

We now specialize to vacuum seeds, $p_0=0$, with a spatially uniform mean pressure $P=C$. This choice supplies the explicit nonzero vacuum endpoint and the nonnegative vacuum magnetic term used in our Chew--Goldberger--Low (CGL)~\cite{cgl} proofs. Our kinetic realization also uses the uniform mean pressure and globally constant $f$. These stability and kinetic results have not been established here for a general pressure-carrying seed. The constant-$f$ action identity remains valid for any seed. We use the exact vacuum fields of Ref.~\cite{sengupta}. They are curl-free and divergence-free, are given in closed form, have nested toroidal surfaces about a regular magnetic axis and nonzero rotational transform, and have no continuous symmetry.

We take $\B_0$ to be a smooth, nonvanishing vacuum field on a toroidal plasma domain, with a regular magnetic axis and nested surfaces labelled by $\psi$:
\begin{equation}
\nabla\times\B_0=0,\qquad\nabla\cdot\B_0=0,\qquad\B_0\cdot\nabla\psi=0.
\label{seed}
\end{equation}
For such a seed we choose a positive smooth function $f(\psi)$ and a spatially uniform pressure $P$. Then
\begin{equation}
\B=f\B_0,\qquad p_\parallel=P+Q,\qquad p_\perp=P-Q,
\qquad Q=\frac{(f^2-1)B_0^2}{2\mu_0}
\label{family}
\end{equation}
solve the static anisotropic equations~\eqref{static} exactly. Here $\PP=p_\perp\I+(p_\parallel-p_\perp)\bb\bb$ and $\bb=\B/B$. For $f=1$ the anisotropy vanishes and the magnetic field is the vacuum seed. With $P=0$ as well this is the vacuum itself, the endpoint $\eta=0$ of the vacuum-connected family given in Methods. The physical reason is a cancellation of stresses:
\begin{equation}
\PP+\frac{B^2}{2\mu_0}\I-\frac{\B\B}{\mu_0}
=P\:\I+\frac{B_0^2}{2\mu_0}\I-\frac{\B_0\B_0}{\mu_0}.
\label{stress}
\end{equation}
The right-hand side is divergence-free. Pressure anisotropy compensates precisely for the change in magnetic stress. Appendix~\ref{app:2} gives the derivation.

Because $f$ is constant along each field line, all field-line trajectories and their rotational transforms are unchanged. This preservation is exact at finite amplitude, with no perturbation expansion. The current is
\begin{equation}
\mathbf j=\frac{f'(\psi)}{\mu_0}\nabla\psi\times\B_0.
\label{current}
\end{equation}
This current is already divergence-free, so no driven parallel-current correction is required. The uniform mean pressure removes that source of the current-solvability problem; it does not impose quasisymmetry (Appendix~\ref{app:2}). Constant $f$ gives a current-free magnetic field supporting finite anisotropic pressure. A surface-dependent $f$ gives a perpendicular plasma current. Both cases retain the same physical magnetic surfaces. No continuous spatial symmetry or quasisymmetry is imposed. In Boozer coordinates the field strength of the sheared configuration has isolated maxima and minima enclosed by closed contours (Fig.~\ref{fig:sheared}d), which quasisymmetry excludes. The Boozer spectrum of the paraboloidal benchmark mixes several helicities (Appendix~\ref{app:3}). A one-parameter family of these equilibria, equation~\eqref{branch} in Methods, switches on the pressure, the anisotropy and the current from the vacuum while the domain, the surfaces and the transform stay fixed.

\subsection*{A nonsymmetric configuration with finite shear}
The $N=1$ ellipsoidal configuration of the vacuum family has nested toroidal surfaces around a planar noncircular magnetic axis (Fig.~\ref{fig:sheared}a,b). It has nonzero magnetic shear such that the rotational transform falls by $7.9\%$, from $\sqrt2/7$ on the axis to $0.186$ at the boundary. Multiplication by a constant or surface-dependent $f$ keeps the field lines and this transform profile. Derivatives with respect to physical flux change only through the flux relabeling, $d\iota/d\psi=(d\iota_0/d\psi_0)/f$. The companions use $P=15$ kPa and $f^2=1.05$ or $1+0.05(1-\rho_0^2)$, the latter shown in Fig.~\ref{fig:sheared}c, and both have $p_\parallel\ge p_\perp$. Started from the projected exact solution, DESC~\cite{dudt,panici} runs with zero net toroidal current reproduce both companions with errors near $10^{-8}$, with the transform and its shear computed as outputs. Appendix~\ref{app:1} gives the parameters and the DESC errors. We mapped the second adiabatic invariant of this configuration (equation~\eqref{action} below) over the labels of nearby trapped wells, for several field-line sections and twelve mirror strengths. None of these 36 maps has an interior minimum or maximum, so local drift confinement is shown on a separate, also sheared, $N=3$ configuration below.

\subsection*{Stability}
The stress identity~\eqref{stress} provides a remarkable organization of the stability results. For a displacement fixed at the boundary, the quadratic energy of every configuration splits into a nonnegative vacuum-seed magnetic term and a local compression and parallel-strain term. In this decomposition the usual pressure-gradient-curvature drive cancels exactly. We use three models. Ideal CGL dynamics gives a sufficient linear criterion for every configuration and a finite-amplitude constrained minimum for $p_\parallel\ge p_\perp$, indicating nonlinear stability. For globally constant $f$, guiding-center kinetic theory gives an electrostatic ground state and a restricted linear electromagnetic criterion. Both kinds of statement are needed. For globally constant $f$, the Kruskal--Oberman quadratic form is bounded above by the CGL form for the same displacement, so CGL stability alone does not imply electromagnetic kinetic stability~\cite{taylorhastie,newcomb}. One pressure inequality, derived first on the kinetic side, governs both.

\emph{Kinetic pressure support.} An exact pressure tensor is not sufficient to ensure equilibria at the kinetic level since the particles must be able to sustain it. For constant $f$, $Q$ is proportional to $B^2$, so both pressures are functions of $B$ alone (Appendix~\ref{app:4}). On a bounded field-strength interval, the guiding-center moment problem can be inverted explicitly. A particularly useful result is that a nonnegative distribution $F(\mathcal E,\mu)$, non-increasing in energy at fixed magnetic moment, exists if and only if
\begin{equation}
p_\perp^2\leq 6Pp_\parallel
\label{bound}
\end{equation}
throughout that interval, assuming positive pressures. The construction follows from an Abel-type moment representation (Methods and appendices). The necessity is the classical Cauchy--Schwarz moment condition of Taylor and Hastie~\cite{taylorhastie}, their Eq.~(3.7), specialized to this pressure law. Our result is an exact statement within the standard guiding-center model, where $\mu$ is an invariant. It is not an exact finite-gyroradius Vlasov realization.

\emph{Linear CGL stability.}  Anisotropic stability and guiding-center pressure support have a long history \cite{kruskaloberman,taylor,HallMcnamara1975}. Earlier work studied the linear stability~\cite{clemente1995stability} and classical particle transport~\cite{clemente1995transport} of anisotropic axisymmetric field-reversed configurations. We take a displacement $\xx$ that vanishes at the boundary and write $c=\nabla\cdot\xx$ and $s=b_i b_j\partial_j\xi_i$. For every equilibrium~\eqref{family}, including surface-dependent $f$, its quadratic potential energy is
\begin{equation}
\delta W=\frac{1}{2\mu_0}\int|\nabla\times(\xx\times\B_0)|^2\,dV+\frac12\int\left[2Pc^2-2(P+Q)cs+(3P+5Q)s^2\right]dV.
\label{energy}
\end{equation}
The first term is the energy of bending and compressing the vacuum seed. The remaining term is a local compression and parallel-strain energy, the quadratic form
\[
2Pc^2-2(P+Q)cs+(3P+5Q)s^2=\begin{pmatrix}c&s\end{pmatrix}\begin{pmatrix}2P&-(P+Q)\\-(P+Q)&3P+5Q\end{pmatrix}\begin{pmatrix}c\\s\end{pmatrix}.
\]
Its matrix has leading entry $2P>0$ and determinant $6Pp_\parallel-p_\perp^2$. Condition~\eqref{bound} therefore makes every term nonnegative. Appendix~\ref{app:5} gives the derivation.

\begin{figure}[tbp]\centering\includegraphics[width=.94\linewidth]{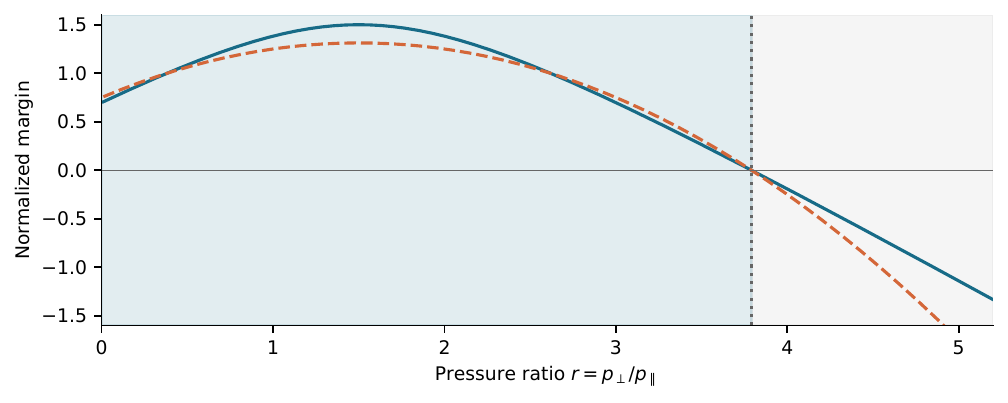}
\caption{\textbf{The same pressure inequality controls two distinct questions.} For $r=p_\perp/p_\parallel$, the normalized CGL compression matrix has diagonal entries $1+r$ and $4-r$ and off-diagonal entries $-1$. The solid blue curve is its minimum eigenvalue. The dashed orange curve is its determinant $3(1+r)-r^2$, the kinetic moment margin, divided by 4. Both vanish at $r=(3+\sqrt{21})/2\approx3.79$ (dotted line). In the blue region the local CGL energy is nonnegative. For the globally constant-$f$ pressure law, an energy-monotone kinetic realization exists if and only if the bound holds throughout the field-strength interval. A value in the grey region leaves the CGL test inconclusive and excludes such a kinetic realization. The CGL statement is sufficient for the entire vacuum-companion family satisfying this bound, and failure of the bound alone does not imply instability. The curves are analytic criteria.}\label{fig:bound}
\end{figure}

The shared inequality is not a coincidence. With $P=(p_\parallel+p_\perp)/2$, the local form in equation~\eqref{energy} can be written $2P(c-s)^2+2p_\perp(c-s)s+3p_\parallel s^2$. For the energy-monotone distributions of a globally constant $f$, its three coefficients are the moments $\sum_s\int(-\partial_{\mathcal E}F_s)(\mu B)^2d^3v=2P$, $\sum_s\int(-\partial_{\mathcal E}F_s)(\mu B)(m_sv_\parallel^2)d^3v=p_\perp$ and $\sum_s\int(-\partial_{\mathcal E}F_s)(m_sv_\parallel^2)^2d^3v=3p_\parallel$ (Appendix~\ref{app:6}). Whenever such a distribution exists, the local CGL energy is a nonnegative sum of squares weighted by $-\partial_{\mathcal E}F_s$. Its determinant $6Pp_\parallel-p_\perp^2$ is the Cauchy--Schwarz moment margin. This is why condition~\eqref{bound} governs both kinetic realization and nonnegative CGL energy (Fig.~\ref{fig:bound}).

This is a sufficient fixed-boundary ideal CGL result, not a necessary criterion. It includes localized ballooning and internal kink stability. With $P=(p_\parallel+p_\perp)/2$ and $r=p_\perp/p_\parallel$, condition~\eqref{bound} reads $r^2-3r-3\le0$, that is $0<r\le(3+\sqrt{21})/2\approx3.79$ (Fig.~\ref{fig:bound}). It holds automatically for $r\le1$, which includes every case with $Q\ge0$, so it restricts only $Q<0$. Methods discusses the equality case and the firehose margin. Although the fluid bound applies to radial $f$, the distribution theorem established here applies to constant $f$.

\emph{Finite-amplitude CGL minimum.} For $0\leq Q<P$, the fluid family has a further finite-amplitude property. Every case minimizes CGL potential energy over smooth ideal material deformations connected to identity and fixed pointwise at the boundary. The conserved constraints are material mass, frozen magnetic flux and the two material invariants $p_\perp/(nB)$ and $p_\parallel B^2/n^3$. The exact finite-deformation identity is
\begin{equation}
E_{\rm CGL}[\mathbf X]-E_{\rm CGL}[\mathrm{id}]
=\frac1{2\mu_0}\int|\B_{0,\mathbf X}-\B_0|^2dV+\int G\,d^3a\geq0,
\label{minimum}
\end{equation}
where $\B_{0,\mathbf X}$ is the frozen image of the vacuum seed and $G\geq0$ is the local compression/strain remainder derived in Appendix~\ref{app:7}, which also compares this minimum with Taylor's relaxation theory~\cite{taylorrelax}, which has found extensive applications in laboratory, astrophysical and solar plasmas. Adding kinetic energy gives a conserved nonnegative relative energy, which is a nonlinear Lyapunov functional for stability. It bounds a perturbed ideal motion of these equilibria, measured in this energy, for as long as that motion remains smooth.

Sampled finite boundary-fixed deformations raised the material CGL energy of the sheared ellipsoidal configuration and of the paraboloidal benchmark, apart from signed values at roundoff level (Appendices~\ref{app:1} and~\ref{app:3}).

\emph{Kinetic stability.} For globally constant $f$ satisfying equation~\eqref{bound}, we choose compatible quasineutral species distributions $F_s(\mathcal E_s,\mu)$ non-increasing in energy. In a fixed magnetic field, these minimize particle energy among phase-space-volume-preserving rearrangements at each $\mu$. With a closed, energy-conserving boundary, nonnegative electrostatic field energy makes the combined state a constrained ground state in the Hamiltonian guiding-center--Poisson model. This applies Gardner's and Helander's framework \cite{gardner,helander}, consistent with the kinetic variational viewpoint of Ref.~\cite{fubhattacharjee}; neither spatial symmetry nor conservation of bounce action is required. For smooth strictly decreasing distributions, a positive quadratic invariant of the linearized fixed-field guiding-center--Poisson model excludes growing electrostatic modes in its norm, and a convex energy--Casimir functional extends this to nonlinear energetic stability (Methods and Appendix~\ref{app:8}).

For globally constant $f$, classical Taylor--Hastie theory~\cite{taylorhastie} supplies a sufficient linear electromagnetic stability argument in the low-frequency small-Larmor-radius guiding-center model~\cite{kruskaloberman}. For these equilibria its coefficients reduce exactly, and its potential energy is the vacuum-seed magnetic energy plus a nonnegative kinetic term (Appendix~\ref{app:6}). Its application requires energy-monotone compatible species, finite moments, a closed supported conservative boundary and well-defined orbit averages. The explicit nonsmooth distributions require a finite quadratic-form or justified limiting interpretation. Kinetic stability for radial $f$ and general nonlinear electromagnetic or finite-Larmor-radius stability are not established here.

\subsection*{Confined drift contours without a symmetry requirement}
Magnetic flux surfaces constrain parallel motion, but particles experiencing magnetic drifts need not remain on them. At fixed energy and magnetic moment, leading bounce-averaged drift trajectories follow level contours of the second adiabatic invariant defined as \cite{HallMcnamara1975,CaryShasharina1997}
\begin{equation}
J(\mathcal E,\mu)=2\int_{\ell_-}^{\ell_+}\sqrt{2m_p(\mathcal E-\mu B)}\,d\ell.
\label{action}
\end{equation}
A nondegenerate definite extremum of this smooth action has nearby closed regular contours, provided the reduced drift coefficient is nonsingular and the same mirror well persists. This local confinement mechanism does not require the action to be constant on every magnetic surface.

\begin{figure}[tbp]\centering\includegraphics[width=\linewidth]{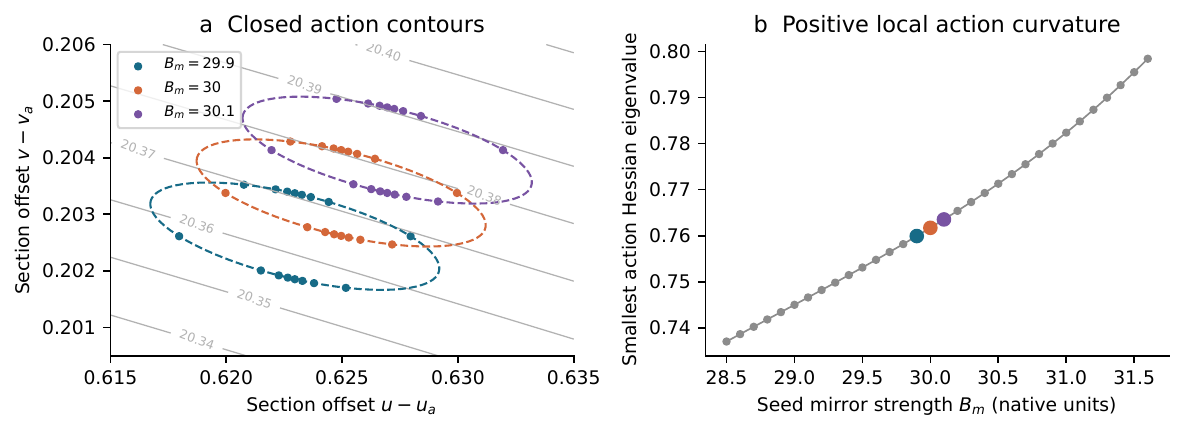}
\caption{\textbf{Local iso-action confinement across magnetic surfaces.} \textbf{a}, Sixteen computed contour roots for each of three seed mirror strengths in the $N=3$ ellipsoidal configuration. The dashed lines are quadratic Hessian approximations, not integrated particle orbits. Grey curves are magnetic surface labels on the section. Coordinates and field strengths use the seed's native normalization. \textbf{b}, Smallest Hessian eigenvalue at the numerical minima for 32 seed mirror strengths. The three coloured points are the mirror strengths of panel a. For constant $f=\sqrt{1.01}$, equation~\eqref{actionmap} transfers the action contours to the anisotropic companion. This configuration, which is also sheared, differs from the $N=1$ configuration of Fig.~\ref{fig:sheared}.}\label{fig:action}
\end{figure}

We now consider the $N=3$ ellipsoidal configuration, which differs from the $N=1$ case shown in Fig.~\ref{fig:sheared}. This configuration is also sheared. Tracing the field lines of the exact seed gives a rotational transform that falls from $\sqrt2/7$ on the axis to between $0.1838$ and $0.1847$ on the surfaces that carry the contours of Fig.~\ref{fig:action}a, about $9\%$ lower (Appendix~\ref{app:9}). Each of 32 mirror strengths from $B_m=28.5$ to $31.6$ gives a numerical action minimum with positive Hessian eigenvalues (Fig.~\ref{fig:action}b), and Fig.~\ref{fig:action}a shows the contours for three of them. Magnetic flux surface labels vary along these contours, demonstrating the distinction between a closed drift contour and omnigeneity. For a globally constant $f$, this confinement structure transfers particularly simply:
\begin{equation}
J_f(\mathcal E,\mu;\psi,\alpha)
=J_0(\mathcal E,f\mu;\psi,\alpha).
\label{actionmap}
\end{equation}
Thus closed contours of the seed map to companion contours at a rescaled magnetic moment. Methods states the scope of this transfer.

Any exact vacuum field with nested magnetic surfaces therefore supplies a family of anisotropic equilibria with the same field lines and rotational transform, and finite shear is retained. Both ellipsoidal configurations shown here are sheared, and the $N=3$ configuration also has local closed action contours. Kinetic stability for surface-dependent rescalings and against general nonlinear electromagnetic perturbations remains open, as does whether such states can be observed in experiments or in nature.

\clearpage
\section*{Methods}
\subsection*{Exact construction and domain}
The seed is restricted to an embedded toroidal domain in which its coordinate map and surface foliation are regular and $B_0>0$. Multiplication by a positive smooth flux function preserves divergence freedom and regularity. The divergence of equation~\eqref{stress} establishes force balance without solving an additional differential equation. Positive pressures require $P>|Q|$ pointwise. The dimensionless anisotropy parameter is \cite{mercier, clemente} $\sigma=\mu_0(p_\parallel-p_\perp)/B^2=1-f^{-2}$. Constant $P$ is essential. The positive pressures remain nonzero at the boundary. These equilibria require boundary support. No plasma--vacuum matching is constructed here.

Following Mercier and Cotsaftis \cite{mercier}, the dimensionless anisotropy parameter $\sigma=\mu_0(p_\parallel-p_\perp)/B^2$ is constant along field lines in this construction, although the pressures themselves generally are not. DESC's anisotropic solver corroborates the constructed equilibria. Rodr\'iguez and Bhattacharjee related quasisymmetry to a more general anisotropic equilibrium formulation~\cite{rodriguez}. Our construction does not allow equilibria with field-line-varying $\sigma$. It therefore cannot resolve the overdetermination of quasisymmetry in the way proposed by Rodr\'iguez and Bhattacharjee.

\subsection*{Zero-firehose-margin class and flowing equilibria}
Exact anisotropic force balance also holds for every divergence-free field when $p_\parallel-p_\perp=B^2/\mu_0$ and $p_\parallel+p_\perp$ is uniform~\cite{lortzspies}. Sato \emph{et al.} used this class for local quasisymmetric fields and for fields without rotational transform in asymmetric toroidal domains~\cite{sato2021,sato2022}. Kaltsas \emph{et al.}~\cite{kaltsas} used it to build toroidally asymmetric equilibria and examined their surfaces by field-line tracing. Some of their parameter choices develop outer islands or stochastic regions. The firehose margin $1-\mu_0(p_\parallel-p_\perp)/B^2$ of that class is exactly zero, so it lies outside our family with finite positive $f$.
Related axisymmetric rotating and cylindrical flowing equilibria have also been constructed~\cite{clemente1994,clemente2009}. Extending the present nonsymmetric toroidal construction to flow would require a separate derivation. We focus on cases without flow, but future extensions of our method to rotating equilibria are possible~\cite{clemente1994,clemente2009}.

\subsection*{Continuous connection to the vacuum field}
To exhibit the vacuum connection explicitly, we choose a bounded smooth $h(\psi)$ and set
\begin{equation}
f_\eta^2=1+\eta h(\psi),\qquad P_\eta=\eta P_*,\qquad
0\leq\eta\leq\eta_*,
\label{branch}
\end{equation}
Here $\eta\ge0$ is the loading parameter, and $\eta=0$ is the vacuum. The function $h(\psi)$ fixes the radial shape of the anisotropy, and $\eta_*$ is any value for which $1+\eta h>0$ throughout, and $P_*$ is the mean pressure per unit $\eta$, so $P_\eta=\eta P_*$. The condition $P_*>\sup|hB_0^2|/(2\mu_0)$ keeps both pressures positive. The pressures and current are then
\begin{equation}
\begin{split}
p_{\parallel,\eta}&=\eta\left(P_*+\frac{hB_0^2}{2\mu_0}\right),\qquad
p_{\perp,\eta}=\eta\left(P_*-\frac{hB_0^2}{2\mu_0}\right),\\
\mathbf j_\eta&=\frac{\eta h'(\psi)}{2\mu_0\sqrt{1+\eta h(\psi)}}\nabla\psi\times\B_0.
\end{split}
\label{branchmoments}
\end{equation}
Both pressures are positive for $\eta>0$. The additional stability restrictions are stated below. On the same fixed toroidal domain,
\begin{equation}
\B_\eta\longrightarrow\B_0\ne0,\qquad
p_{\parallel,\eta},p_{\perp,\eta}\longrightarrow0,\qquad
\mathbf j_\eta\longrightarrow0\qquad(\eta\longrightarrow0).
\label{vacuumlimit}
\end{equation}
For smooth seed data this convergence is smooth. No surface or axis degenerates in the limit. If $h$ is not a constant, every configuration with $\eta>0$ carries plasma current. The rotational transform and physical field-line trajectories are unchanged for every $\eta$, not merely to leading order.

This regular vacuum limit distinguishes the family from finite-pressure constructions without such an endpoint. The pressure load is supported by anisotropic stress with spatially uniform arithmetic mean $P_\eta=(p_{\parallel,\eta}+p_{\perp,\eta})/2$; an arbitrary scalar-pressure gradient cannot be added to the cancellation. At an isotropic case $Q=0$, equation~\eqref{family} returns $f=1$ and uniform pressure. Each value of $\eta$ gives an equilibrium with its boundary support. The family is not an ideal dynamical trajectory from one value to another.

\subsection*{Guiding-center moment inversion}
For $\mathcal E=m_pv_\parallel^2/2+\mu B$, we take $F=w(\mu)\widehat G(\mathcal E/\mu)$, where $w\geq0$ has finite positive $\mu^{1/2}$ and $\mu^{3/2}$ moments, as an exponential does, and $\widehat G(y)=\int_y^\infty\nu(b)\,db$, with $\nu$ a nonnegative measure. With
$C_w=4\pi\sqrt{2}\,m_p^{-3/2}\int_0^\infty w(\mu)\mu^{3/2}d\mu$, its moments are
\begin{align}
p_\parallel&=\tfrac23 C_w B\int\nu(b)(b-B)_+^{3/2}db,\\
p_\perp&=C_w B^2\int\nu(b)(b-B)_+^{1/2}db.
\end{align}
Cauchy--Schwarz applied to these moments and their derivatives gives equation~\eqref{bound}. Conversely, a positive atom above $B_{\max}$ plus a nonnegative density on $[B_{\min},B_{\max}]$ reproduces the target pressure law whenever the bound holds. The explicit inversion is in Appendix~\ref{app:4}. More general energy-monotone distributions can be integrated over $\mu$ to obtain the same positive-measure representation. In terms of $R_\mu=(\int w\mu^{1/2}d\mu)/(\int w\mu^{3/2}d\mu)$, the physical moments obey $n=R_\mu p_\perp/B$, $T_\perp=B/R_\mu$ and $T_\parallel=Bp_\parallel/(R_\mu p_\perp)$ in energy units. They generally vary with $B$. Appendix~\ref{app:6} gives a compatible ion--electron normalization with pointwise neutrality. It also shows that for constant $f$ the guiding-center and magnetization currents cancel, as the curl-free field requires.

\subsection*{Pressure bound and firehose margin}
At equality the local form is semidefinite. The bound is sufficient for nonnegative CGL energy because the nonnegative seed magnetic term was not used. It is not a necessary mirror threshold. The ideal firehose margin is $1-\mu_0(p_\parallel-p_\perp)/B^2=f^{-2}>0$ for every positive $f$. For $f<1$ the perpendicular pressure is larger, so condition~\eqref{bound} must also be checked.

\subsection*{Electrostatic invariant}
For smooth strictly decreasing distributions the linearized equations of the fixed-field, $\mu$-conserving guiding-center--Poisson model with a conservative boundary conserve the quadratic invariant derived in Appendix~\ref{app:8} (for energy--Casimir background, see Ref.~\cite{fubhattacharjee} and references therein)
\begin{equation}
\mathcal W_{\rm es}=\frac12\sum_s\int\frac{|\delta f_s|^2}{-\partial_{\mathcal E}F_s}d\Gamma_s
+\frac{\epsilon_0}{2}\int|\nabla\delta\phi|^2dV.
\label{esenergy}
\end{equation}
The kinetic and electrostatic parts change at equal and opposite rates. At the energy--Casimir level this corresponds to $C_s''(F_s,\mu)=1/(-\partial_{\mathcal E}F_s)$~\cite{holm1985}. This positive invariant excludes growing modes in its norm. A convex energy--Casimir functional extends it to nonlinear energetic stability. Our explicit distributions can have plateaus and steps: their ground-state statement uses rearrangement and allows neutral rearrangements (Appendix~\ref{app:8}). The theorem does not cover electromagnetic perturbations or processes that break conservation of $\mu$.

\subsection*{Ideal constraints and second variation}
For a material deformation $\mathsf F$, we write $\mathcal J=\det\mathsf F$ and $\ell=|\mathsf F\bb|$. Frozen flux gives $\B'=\mathsf F\B/\mathcal J$; mass conservation gives $n'=n/\mathcal J$. Conservation of $p_\perp/(nB)$ and $p_\parallel B^2/n^3$ gives the CGL material energy
\begin{equation}
U(\mathsf F)=\frac{B^2\ell^2}{2\mu_0\mathcal J}
+\frac{p_\perp\ell}{\mathcal J}+\frac{p_\parallel}{2\ell^2}.
\end{equation}
Its first variation gives force balance, the anisotropic counterpart of an ideal constrained equilibrium formulation~\cite{kruskal}. Standard variational background is in Refs.~\cite{newcomb,webb}. Expanding $\mathsf F=\I+\epsilon\nabla\xx$ to second order, subtracting the vacuum-seed magnetic energy and using $\xx=0$ on the boundary yields equation~\eqref{energy}. The integrated residual $P[\mathrm{tr}(\nabla\xx)^2-(\nabla\cdot\xx)^2]/2$ vanishes by integration by parts. Each constructed state is therefore an equilibrium of the CGL energy under the ideal constraints. This does not show that it is the only such equilibrium, or that an ideal motion can carry the plasma from one configuration of the family to another.

\subsection*{Action calculations}
The action example uses the ellipsoidal seed with $k=0.05$, $N=3$, $\kappa=1.5$, phase $K(1-k^2)$ and $u_a=v_a\approx1.662556$. Its axis transform is $\sqrt2/7$. Numerical quadrature evaluates $j=\int\sqrt{B_m-B}\,d\ell$, with $J=2\sqrt{2m_p\mu}\,j$. At each of 32 values $B_m=28.5,28.6,\dots,31.6$, action minimization is followed by Hessian estimation and sixteen radial contour-root solves. Closed contours represent leading bounce-averaged drift confinement only while mirror points remain simple and the drift reduction is regular. The contours in Fig.~\ref{fig:action} are not direct integrations of guiding-center or finite-gyroradius orbits. For a surface-dependent $f$, equation~\eqref{actionmap} still holds on each field line, but the rescaled moment $f\mu$ differs between the surfaces that a drift contour crosses. A contour of $J_f$ then need not be a contour of $J_0$ at a single magnetic moment, so the seed's closed contours do not carry over directly. Appendix~\ref{app:9} gives the contour data and the rotational transform.

\subsection*{Configurations treated in the appendices}
Appendix~\ref{app:1} gives the parameters, the DESC errors, the finite-deformation tests and the action maps of the sheared configuration of Fig.~\ref{fig:sheared}. An exact paraboloidal vacuum configuration with exactly constant rotational transform, with constant and surface-dependent $f$, serves as a shearless benchmark. Appendix~\ref{app:3} gives its geometry, Boozer harmonics, DESC results and finite-deformation tests of the energy minimum.

\subsection*{Use of AI tools}
Under the authors' direction, Anthropic Claude Code and OpenAI Codex assisted with analytical derivations, symbolic and numerical checks, equilibrium-code comparisons, scientific plotting code, literature searches and manuscript drafting and editing. The authors set the research questions, selected the examples and reviewed intermediate results and drafts. Checks included analytical identities, symbolic calculations, convergence tests and comparisons between separate agent sessions. The authors retain responsibility for the scientific claims and the final manuscript.

\section*{Data availability}
The data that support the figures and the validation results, including the DESC comparison data, the finite-deformation results and single-mode minimization checks, the action-curvature scan and the field-line calculation of the rotational transform of the $N=3$ configuration, are provided in a source archive (source\_archive\_v12.5.zip) that is available on request and will be deposited in a public repository with a persistent identifier before publication. Full numerical equilibrium files are retained in a separate DESC reproducibility package and are available from the corresponding author on request.

\section*{Code availability}
The scripts that evaluate the exact fields, run the finite-deformation tests, trace the field lines, drive the DESC calculations and generate the figures are contained in the same source archive, which is available on request and will be deposited with the data. DESC is publicly available~\cite{dudt,panici}.

\section*{Acknowledgements}
This research was supported by the Simons Foundation/SFARI (Grant No.\ 560651, AB) and the US Department of Energy
(Award Nos.\ DE-SC0024548 and DE-AC02-09CH11466).

\section*{Correspondence}
Correspondence and requests for materials should be addressed to \textbf{amitava@princeton.edu}.

\clearpage
\appendix
\renewcommand{\thesection}{\Alph{section}}
\counterwithin{figure}{section}
\counterwithin{equation}{section}
\renewcommand{\thefigure}{\thesection\arabic{figure}}
\renewcommand{\theequation}{\thesection\arabic{equation}}
\begin{center}\Large\bfseries Appendices\end{center}

\section{The sheared ellipsoidal configuration of \texorpdfstring{Fig.~\ref{fig:sheared}}{Fig. 1}}\label{app:1}
\begin{figure}[ht]\centering
\includegraphics[width=.94\linewidth]{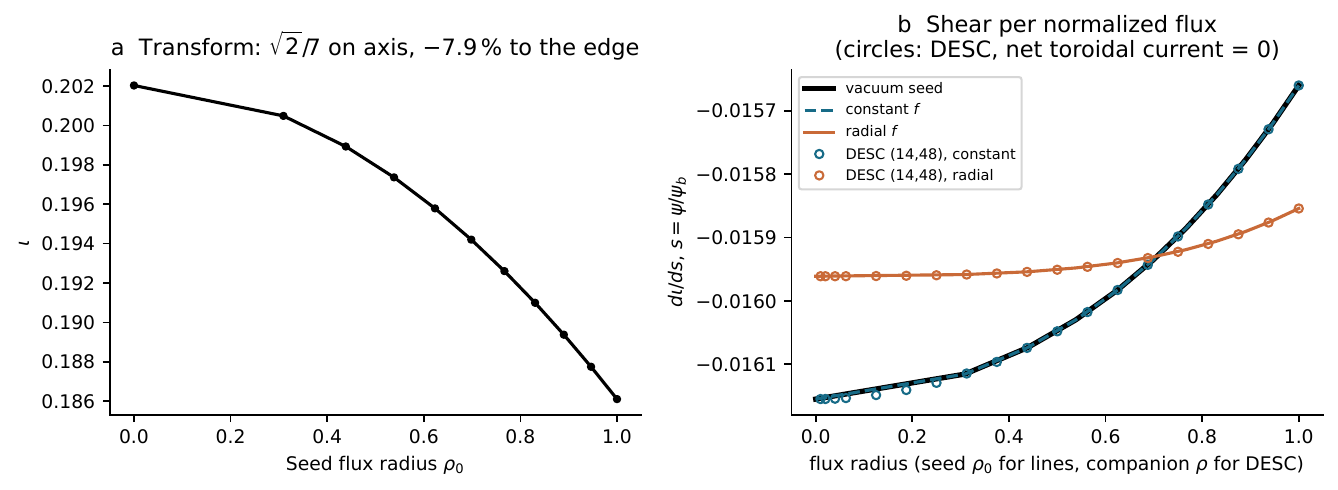}
\caption{\textbf{Rotational transform and shear of the configuration of Fig.~\ref{fig:sheared}.} The exact transform falls by $7.9\%$ from $\sqrt2/7$ on the axis. Circles are DESC values at $(L,M,N)=(14,14,48)$ with zero net toroidal current. The rescaling leaves the transform unchanged, and derivatives with respect to physical flux include the flux relabeling.}\label{sfig:sheared}
\end{figure}

\textbf{Parameters.} The configuration has $(N,M,G)=(1,2,2)$, $k=k_g=3-2\sqrt2$, $\kappa=1/2$, axis transform $\sqrt2/7$ and boundary surface-label excess $0.1001112822$. Native lengths are read as metres, and the field is scaled by $0.012033783$ T per native unit, so the maximum seed field is about $0.616$ T. The companions use $P=15$ kPa and $f^2=1.05$ or $1+0.05(1-\rho_0^2)$. The transform falls from $0.2020305$ on the axis to $0.1861083$ at the boundary, and $d\iota/ds$ with $s=\psi/\psi_b$ runs from about $-0.0162$ to $-0.0157$. Both pressures are positive. The local $\beta=2\mu_0P/B^2$ ranges from $0.094$ to $5.09$ for constant $f$ and from $0.099$ to $5.35$ for radial $f$.

\textbf{DESC.} Each run starts from the exact solution projected on DESC's basis (DESC 0.17.1, two field periods). The net toroidal current is set to zero, so the transform is an output. At $(L,M,N)=(14,14,48)$ the maximum relative field error against the exact solution is $1.1\times10^{-8}$ for both companions, $|\Delta\iota|\le2.9\times10^{-11}$, and the relative error in $d\iota/d\psi$ is at most $2.6\times10^{-8}$. At $(12,12,40)$ the field error is $9.6\times10^{-8}$, the same as for a vacuum run.

\textbf{Energy and action maps.} Finite boundary-fixed deformations raise the material CGL energy of this configuration, as for the benchmark of Appendix~\ref{app:3}. A mixed deformation raises it by about $937$ J, or $0.042\,PV$. For this configuration, none of 36 maps of the second adiabatic invariant over the labels of nearby trapped wells has an interior minimum or maximum. The maps use well-centre and displaced sections and twelve mirror levels from $0.10$ to $0.40$ T.

\section{Exact equilibrium and current}\label{app:2}
We write the total stress as $\mathsf T=\mathsf P+B^2\mathsf I/(2\mu_0)-\B\B/\mu_0$. For $\B=f(\psi)\B_0$, $p_\parallel=P+Q$, $p_\perp=P-Q$ and $Q=(f^2-1)B_0^2/(2\mu_0)$, the identity $(p_\parallel-p_\perp)\bb\bb=(f^2-1)\B_0\B_0/\mu_0$ gives
\begin{equation}
\mathsf T=P\mathsf I+\frac{B_0^2}{2\mu_0}\mathsf I-\frac{\B_0\B_0}{\mu_0}.
\end{equation}
For uniform $P$ and a curl-free, divergence-free seed, $\nabla\cdot\mathsf T=0$. Also $\nabla\cdot(f\B_0)=0$, because $f$ is constant on field lines. Smoothness at the axis requires $f$ to be a smooth function of a regular surface label. In the notation of Fu and Bhattacharjee (Ref.~\cite{fubhattacharjee}), the family has $p(\psi)=P$ and $K(\psi)=(1-f^{-2})/(2\mu_0)$, so $p_{\parallel,\perp}=P\pm KB^2$.

Ampère's law gives
\begin{equation}
\mathbf j=\frac{f'(\psi)}{\mu_0}\nabla\psi\times\B_0,\qquad \mathbf j\cdot\B=\mathbf j\cdot\nabla\psi=0.
\end{equation}
Since $\nabla\times\B_0=0$, $\nabla\cdot(\nabla\psi\times\B_0)=0$, and differentiating $f'(\psi)$ adds a multiple of $\nabla\psi\cdot(\nabla\psi\times\B_0)=0$. Hence $\nabla\cdot\mathbf j_\perp=0$ and $\B\cdot\nabla(j_\parallel/B)=0$. The family has $j_\parallel=0$ without solving a magnetic differential equation, and no quasisymmetry condition enters.

\section{The shearless paraboloidal benchmark}\label{app:3}
\begin{figure}[htp]\centering\includegraphics[width=\linewidth]{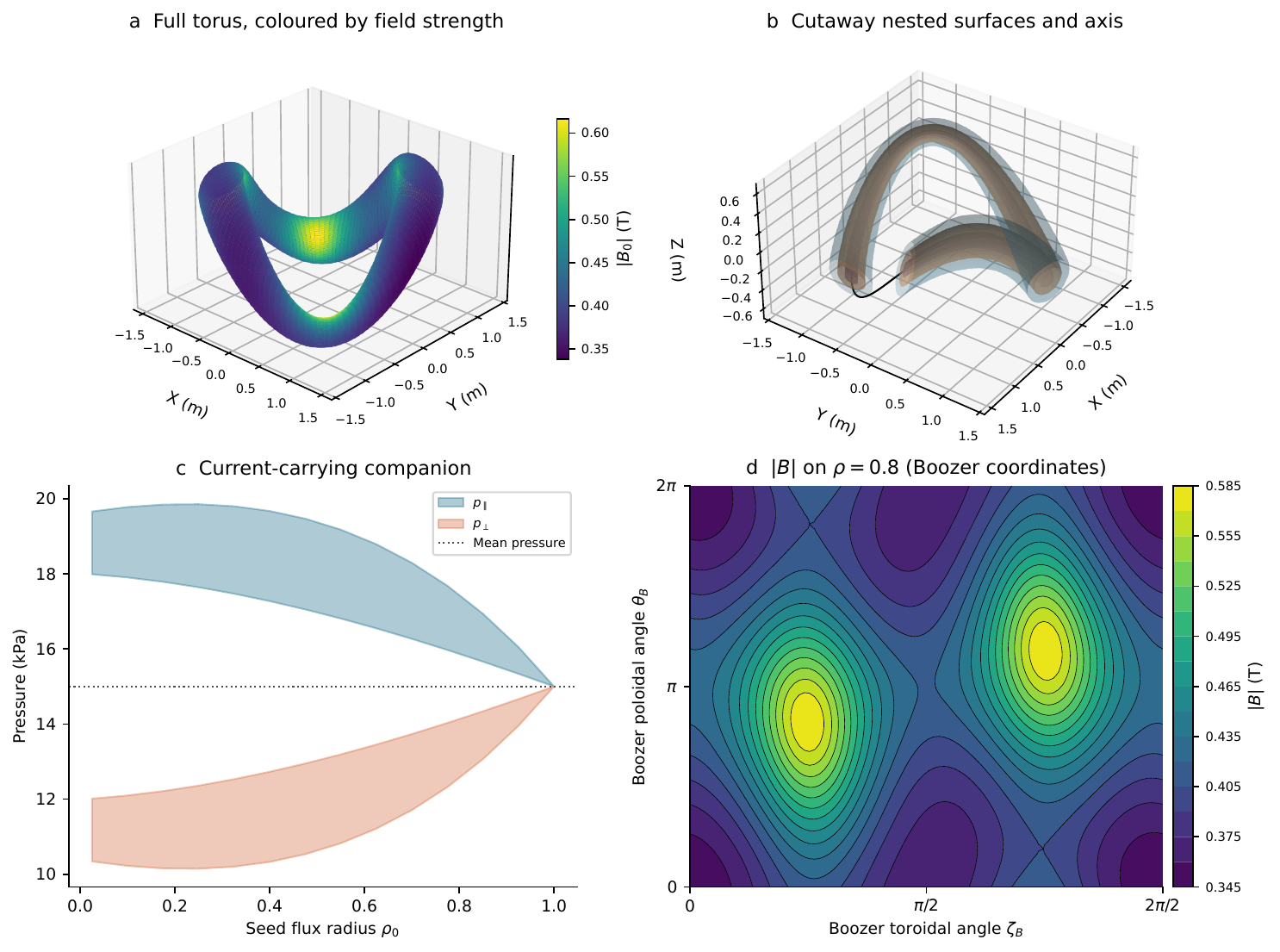}
\caption{\textbf{Shearless paraboloidal benchmark.} \textbf{a}, Outer surface of the seed, coloured by the exact field strength. \textbf{b}, Cutaway with two nested interior surfaces and the nonplanar magnetic axis (black). \textbf{c}, Pressure envelopes of the current-carrying companion $f^2=1+0.05(1-\rho_0^2)$ with $P=15$ kPa, sampled over each surface. \textbf{d}, Contours of the seed field strength $|B|$ in Boozer coordinates at $\rho=0.8$, over one field period.}\label{sfig:geometry}
\end{figure}

\textbf{Geometry.} The position map is
\begin{equation}
\mathbf x=(\sinh u\cos z-\sinh v\sin z,\ \sinh u\sin z-\sinh v\cos z,\ \sinh u\sinh v+\tfrac12\sin2z),
\end{equation}
with $k=0.5$, $u_a=1$, $v_a=0$, $\kappa=2K(1-k^2)/\pi$, zero profile phase and magnetic scale $0.2$ in SI units. The surface label runs from $1$ on the axis to $1.02$. The transform is the constant $0.1103880277$, the toroidal flux is $0.0582526266$ Wb, and the axis, with $Z_a=\tfrac12\sin2z$, is not planar.

\textbf{Boozer harmonics.} At $\rho=0.8$ the Boozer spectrum $B=\sum_{m,n}B^c_{mn}\cos(m\theta_B-n\zeta_B)+\dots$, with $n$ the full-torus mode number, has $B_{00}=0.4288$ T. The leading amplitudes $|B^c_{mn}|/B_{00}$ are $0.117$ for $(m,n)=(1,0)$, $0.111$ for $(0,4)$, $0.054$ for $(1,-2)$, $0.036$ for $(1,2)$, $0.028$ for $(1,-4)$ and $0.018$ for $(1,4)$. Several helicities contribute, so the field is not quasisymmetric on this surface.

\begin{figure}[htbp]\centering\includegraphics[width=\linewidth]{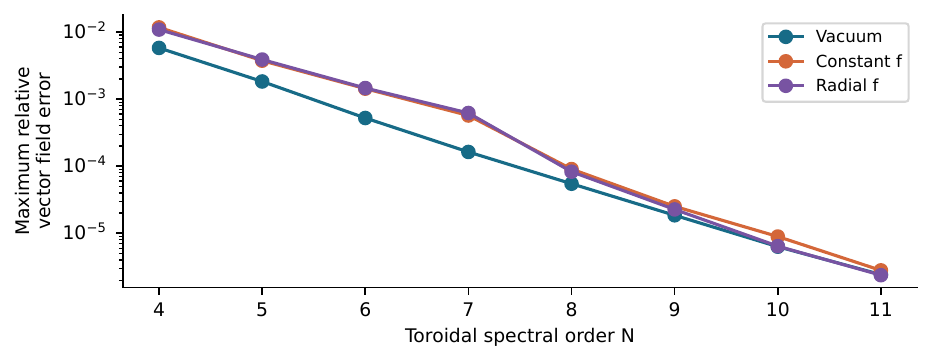}
\caption{\textbf{DESC equilibria against the exact paraboloidal solution.} Maximum relative vector-field error at common physical points on an independent validation grid, at eight resolutions from $(L,M,N)=(6,4,4)$ to $(11,9,11)$. The transform is prescribed.}\label{sfig:desc}
\end{figure}

\textbf{DESC.} DESC 0.17.3 with two field periods solves the vacuum and the configurations of equation~\eqref{branch} with $h=1$ (constant $f$) or $h=1-\rho_0^2$ (radial $f$), $P_*=3\times10^5$ Pa and $\eta=0.05$, so $P=15$ kPa. The transform is prescribed, and the three-dimensional $p_\perp$ is fitted in Fourier--Zernike modes. The displayed runs start from the exact geometry fitted on DESC's basis with $\lambda=0$. From $(L,M,N)=(6,4,4)$ to $(11,9,11)$ the maximum relative field error against the exact solution falls from at most $1.2\times10^{-2}$ to $2.4\times10^{-6}$ for the vacuum, $2.8\times10^{-6}$ for constant $f$ and $2.4\times10^{-6}$ for radial $f$ (Fig.~\ref{sfig:desc}). At $(11,9,11)$ the normalized force residuals are at most $2.3\times10^{-6}$. Vacuum points at $N\ge7$ reach the 100-iteration limit. The $(12,10,12)$ runs are excluded because the six-radius initial fit is rank deficient and gives an invalid starting geometry. For radial $f$ at $(8,6,8)$ the computed current agrees with equation~\eqref{current} to $0.46\%$ RMS.

\begin{figure}[htbp]\centering\includegraphics[width=\linewidth]{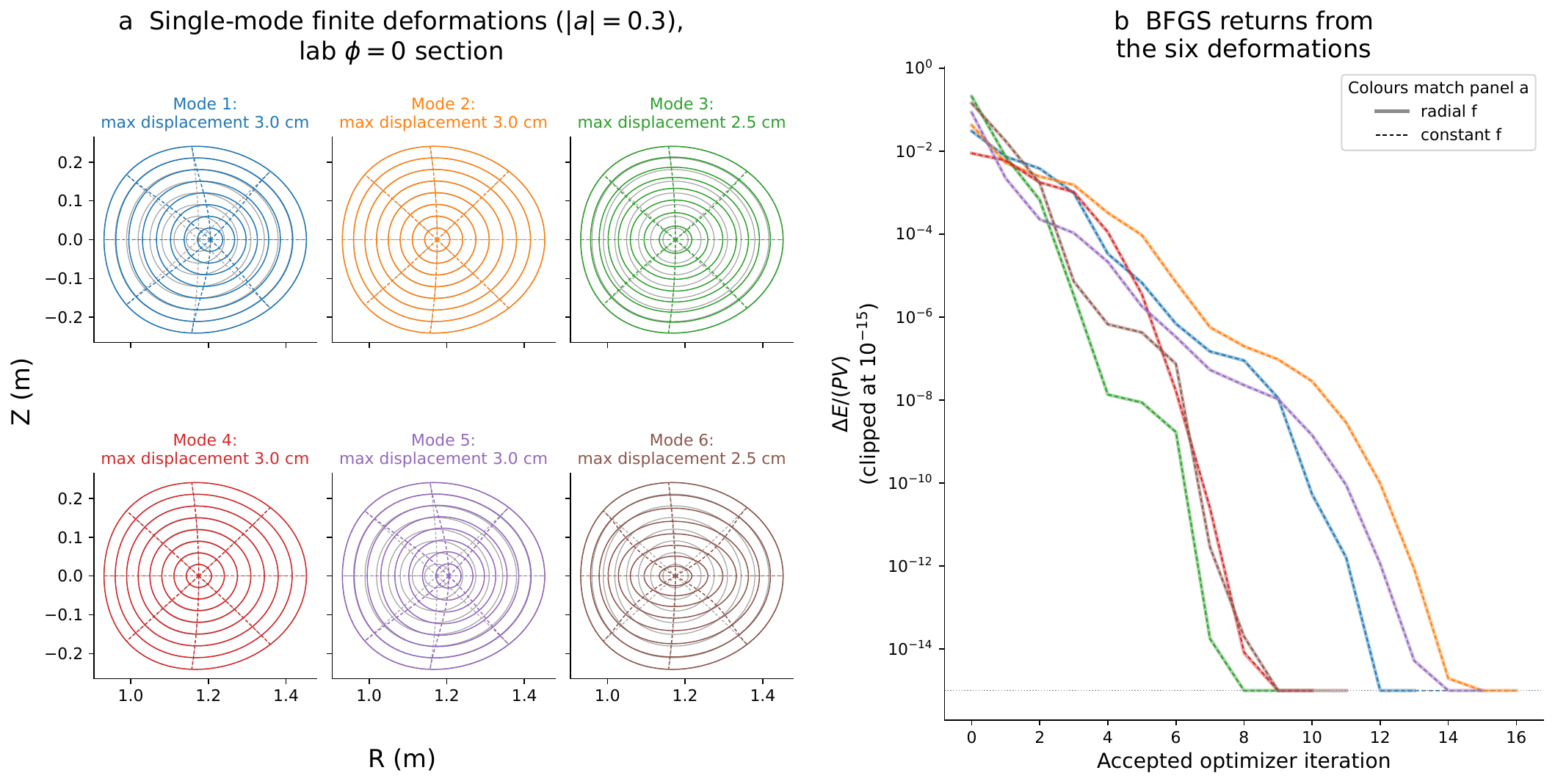}
\caption{\textbf{Finite deformations raise the CGL energy.} \textbf{a}, Sections at $\phi=0$ of the six single-mode deformations $\xx_1$ to $\xx_6$ at amplitude $0.3$. Modes 2 and 4 vary as $\sin z$ and $\sin2z$ and are nearly invisible in this section. \textbf{b}, Energy during BFGS minimization from these six starts on both companions. All twelve runs return to the unperturbed reference. Energies are normalized by $PV=18.05$ kJ and clipped at $10^{-15}$ for display. The initial deformations displace the plasma by up to $3.0$ cm.}\label{sfig:minimum}
\end{figure}

\textbf{Finite deformations.} We evaluate the material CGL energy, equation~\eqref{U} below, directly after boundary-fixed deformations $\mathbf X=\mathbf x+\sum_ia_i\xx_i$ of both companions, with the physical volume $V=1.2033$ m$^3$. Flux, mass and both CGL invariants are advected, and the pressure law is not imposed again. With $\mathbf x_a(z)=\mathbf x(1,0,z)$, $\mathbf d=\mathbf x-\mathbf x_a$, $d_R=\mathbf d\cdot\mathbf e_R$, $d_Z=\mathbf d\cdot\mathbf e_Z$, the envelope $w=[1-(\psi-1)/0.02]^2$ and $L=0.1$ m, the six modes are
\begin{equation}
\begin{gathered}
\xx_1=wL\cos z\,\mathbf e_R,\quad
\xx_2=wL\sin z\,\mathbf e_Z,\quad
\xx_3=w(d_R\mathbf e_R+d_Z\mathbf e_Z),\\
\xx_4=wL\sin2z\,\mathbf e_\phi,\quad
\xx_5=wL(\cos3z\,\mathbf e_R+\sin3z\,\mathbf e_Z),\quad
\xx_6=w(d_R\mathbf e_R-d_Z\mathbf e_Z).
\end{gathered}
\end{equation}
For $\mathbf a=(0.15,-0.12,0.10,0.10,-0.08,0.08)$, the energy rises by $0.05802\,PV$ for both companions. This agrees with the identity~\eqref{globalmin} to $2\times10^{-10}$ on the finest grid. Forty random mixed deformations give $\Delta E/(PV)$ between $0.0049$ and $0.113$. Both signs of every mode raise the energy for coefficients from $10^{-4}$ to $0.3$. For the energy minimizations we start from each mode alone at amplitude $0.3$. These starts displace the plasma by up to $3.0$ cm, keep the minimum Jacobian above $0.63$ and raise the energy by $0.0089\,PV$ (mode 4) to $0.209\,PV$ (mode 3). BFGS minimization returns all twelve runs, six on each companion, to the reference's numerical neighbourhood within 9 to 16 iterations, with coefficient norms below $1.4\times10^{-9}$ and $|\Delta E|/(PV)<5\times10^{-17}$ on the finest grid (Fig.~\ref{sfig:minimum}). The Hessian in this subspace has eigenvalues from $0.196$ to $5.19$.

\section{Energy-monotone kinetic realization}\label{app:4}
We take constant $f$ and $K=(1-f^{-2})/(2\mu_0)$, so that $p_\parallel=P+KB^2$ and $p_\perp=P-KB^2$ on $0<B_{\min}\leq B\leq B_{\max}$, in the guiding-center model with no electric potential. For $F(\mathcal E,\mu)\ge0$ non-increasing in $\mathcal E$, a superposition of energy steps gives
\begin{align}
p_\parallel&=\tfrac23 C_w B\int\nu(b)(b-B)_+^{3/2}db,\label{mom1}\\
p_\perp&=C_w B^2\int\nu(b)(b-B)_+^{1/2}db,\label{mom2}
\end{align}
with a nonnegative measure $\nu$. One realization is $F=w(\mu)\int_{\mathcal E/\mu}^\infty\nu(b)db$ with $w\geq0$ and $C_w=4\pi\sqrt2\,m_p^{-3/2}\int_0^\infty w\mu^{3/2}d\mu$. A general $\mu$-dependent mixture gives the same representation after integration over $\mu$.

\textbf{Necessity.} With $\phi=p_\parallel/B$, the representation gives $\phi'=-p_\perp/B^2$ and $\phi''=(2p_\perp-B\partial_Bp_\perp)/B^3$. Cauchy--Schwarz gives $(\phi')^2\leq3\phi\phi''$, that is $p_\perp^2\leq3p_\parallel(2p_\perp-B\partial_Bp_\perp)$. For the target law the bracket is $2P$, which gives $p_\perp^2\leq6Pp_\parallel$.

\textbf{Sufficiency.} Hats denote values at $B_{\max}$. We set
\begin{equation}
\Psi(B)=\frac{3(P+KB^2)}{2C_wB},\quad
d=\frac{3B_{\max}\widehat p_\parallel}{2\widehat p_\perp},\quad
a=\Psi(B_{\max})d^{-3/2},
\end{equation}
$x=B_{\max}-B$, $r(x)=\Psi(B_{\max}-x)-a(d+x)^{3/2}$ and
\begin{equation}
\rho(x)=\frac{r''(0)x^{-1/2}+\int_0^x r'''(t)(x-t)^{-1/2}dt}{\sqrt\pi\,\Gamma(5/2)}.
\end{equation}
The measure is the density $\rho(B_{\max}-b)$ on $[B_{\min},B_{\max}]$ plus an atom $a\delta(b-B_{\max}-d)$. The choices of $d$ and $a$ give $r(0)=r'(0)=0$, and
\begin{equation}
r'''(t)=\frac{9P}{C_w(B_{\max}-t)^4}+\frac{3a}{8(d+t)^{3/2}}>0,\qquad
r''(0)=\frac{3P}{C_wB_{\max}^3}\left(1-\frac{\widehat p_\perp^2}{6P\widehat p_\parallel}\right).
\end{equation}
The bound therefore makes $\rho\ge0$. Fractional integration gives $\int\nu(b)(b-B)_+^{3/2}db=\Psi(B)$, which is equation~\eqref{mom1}, and differentiation gives equation~\eqref{mom2}. For $K<0$, $p_\perp/p_\parallel$ increases with $B$, so the bound at $B_{\max}$ implies it on the whole interval. For $K\geq0$ the bound holds automatically. A quadrature check at $f=\sqrt{1.01}$ reproduces the analytic moments to $4\times10^{-15}$.

\section{CGL quadratic energy}\label{app:5}
Particle number, frozen flux and the two CGL invariants $p_\perp/(nB)$ and $p_\parallel B^2/n^3$ are conserved. For a deformation gradient $\mathsf F$ with $\mathcal J=\det\mathsf F$ and $\ell=|\mathsf F\bb|$,
\begin{equation}
n'=\frac n{\mathcal J},\quad B'=\frac{B\ell}{\mathcal J},\quad
p_\perp'=\frac{p_\perp\ell}{\mathcal J^2},\quad
p_\parallel'=\frac{p_\parallel}{\mathcal J\ell^2}.
\end{equation}
The internal energy density is $p_\perp'+p_\parallel'/2$, and the material potential energy per initial volume is
\begin{equation}
U=\frac{B^2\ell^2}{2\mu_0\mathcal J}+\frac{p_\perp\ell}{\mathcal J}+\frac{p_\parallel}{2\ell^2}.
\label{U}
\end{equation}
Its first variation is the virtual work of the total stress, so the stress identity of Appendix~\ref{app:2} gives stationarity.

For $\mathsf F=\mathsf I+\epsilon A$ with $A_{ij}=\partial_j\xi_i$, we write $c=\mathrm{tr}A$, $s=b_iA_{ij}b_j$ and $t=|A\bb|^2$. Then
\begin{equation}
\mathcal J=1+\epsilon c+\frac{\epsilon^2}{2}(c^2-\mathrm{tr}A^2)+O(\epsilon^3),\qquad
\ell^2=1+2\epsilon s+\epsilon^2t.
\end{equation}
We subtract the magnetic energy of the seed frozen through the same deformation, $B_0^2\ell^2/(2\mu_0\mathcal J)$. With $p_{\parallel,\perp}=P\pm Q$ and $Q=(B^2-B_0^2)/(2\mu_0)$, the second-order coefficient is
\begin{equation}
\frac12\left[2Pc^2-2(P+Q)cs+(3P+5Q)s^2\right]+\frac P2(\mathrm{tr}A^2-c^2).
\end{equation}
For constant $P$ and $\xx=0$ on the boundary, the last term integrates to zero. The seed term gives $\int|\nabla\times(\xx\times\B_0)|^2dV/(2\mu_0)$, which is equation~\eqref{energy}. The local matrix
\begin{equation}
\mathsf M=\begin{pmatrix}2P&-(P+Q)\\-(P+Q)&3P+5Q\end{pmatrix},\qquad
\det\mathsf M=5P^2+8PQ-Q^2=6Pp_\parallel-p_\perp^2,
\end{equation}
is nonnegative under the bound, so $\delta W\geq0$. Positive mass density then excludes exponentially growing ideal CGL normal modes in this boundary class.

\section{Kinetic closure and Taylor--Hastie stability for constant \texorpdfstring{$f$}{f}}\label{app:6}
We use the low-frequency small-Larmor-radius guiding-center energy principle of Kruskal and Oberman (Ref.~\cite{kruskaloberman}) and Taylor and Hastie (Ref.~\cite{taylorhastie}), with globally constant $f$, no equilibrium electric potential, $\xx=0$ on the boundary and closed particle dynamics that conserve energy, $\mu$ and phase-space volume.

\textbf{Coefficients.} Parallel force balance gives
\begin{equation}
B\partial_Bp_\parallel=p_\parallel-p_\perp,\quad
B\partial_Bp_\perp-2p_\perp=-2P,\quad
1+\frac{\mu_0(p_\perp-p_\parallel)}{B^2}=1+\frac{\mu_0}{B}\partial_Bp_\perp=f^{-2}.
\end{equation}
These are exact, not low-beta approximations.

\textbf{Moment identity.} With $c=\nabla\cdot\xx$, $h=b_ib_j\partial_j\xi_i$ and $A_s=m_sv_\parallel^2h+\mu B(c-h)$, integration by parts in $v_\parallel$ and summation over species give
\begin{equation}
\sum_s\int(-F_{s,\mathcal E})(\mu B)^2d^3v=2P,\quad
\sum_s\int(-F_{s,\mathcal E})(\mu B)(m_sv_\parallel^2)d^3v=p_\perp,\quad
\sum_s\int(-F_{s,\mathcal E})(m_sv_\parallel^2)^2d^3v=3p_\parallel.
\end{equation}
The Gram determinant is $6Pp_\parallel-p_\perp^2$, and
\begin{equation}
2P(c-h)^2+2p_\perp(c-h)h+3p_\parallel h^2=\sum_s\int(-F_{s,\mathcal E})A_s^2d^3v.
\end{equation}
The Kruskal--Oberman energy replaces $A_s^2$ by the square of its orbit average,
\begin{equation}
\delta W_{\rm KO}=\frac1{2\mu_0}\int|\nabla\times(\xx\times\B_0)|^2dV
+\frac12\sum_s\int(-F_{s,\mathcal E})\langle A_s\rangle^2d\Gamma_s\geq0,
\end{equation}
and $\delta W_{\rm CGL}-\delta W_{\rm KO}=\frac12\sum_s\int(-F_{s,\mathcal E})(A_s-\langle A_s\rangle)^2d\Gamma_s\geq0$. For the step distributions of Appendix~\ref{app:4}, $-F_{s,\mathcal E}$ is a nonnegative measure and enters linearly, so these integrals hold in the Stieltjes sense when they are finite. As stated in the main text, using them as a dynamical statement requires a finite quadratic-form or justified limiting interpretation.

\textbf{Current and neutrality.} With $\boldsymbol\kappa_B=(\bb\cdot\nabla)\bb$, an even distribution in $v_\parallel$ gives
\begin{equation}
\mathbf J_{\rm gc}=\frac{p_\perp}{B^2}\bb\times\nabla B+\frac{p_\parallel}{B}\bb\times\boldsymbol\kappa_B,\quad
\mathbf M=-\frac{p_\perp}{B}\bb,\quad
\mathbf J_{\rm gc}+\nabla\times\mathbf M=\frac1B\bb\times[\nabla p_\perp+(p_\parallel-p_\perp)\boldsymbol\kappa_B]=0,
\end{equation}
using the vacuum curvature $\boldsymbol\kappa_B=\nabla_\perp\ln B$ and $p_\perp=P-KB^2$. This agrees with $\nabla\times\B=0$. For singly charged ions and electrons with the same step measure in $\mathcal E/\mu$ and the same moment ratio $R_\mu$, the amplitudes can be chosen so that $P_i=P_e=P/2$ and $K_i=K_e=K/2$. Then $n_s=R_\mu p_{\perp s}/B$ gives $n_i=n_e$ at every point.

\section{Finite-deformation CGL minimum}\label{app:7}
We take smooth orientation-preserving material maps $\mathbf X(\mathbf a)$ of the domain, connected to the identity through maps fixed at the boundary, with $\mathsf F=\partial\mathbf X/\partial\mathbf a$, $\mathcal J>0$ and $\ell>0$. Mass, flux and the two CGL invariants are advected. For uniform $P>0$ and $0\leq Q<P$, which allows radial $f$, the reference state minimizes the potential energy~\eqref{U}.

\textbf{Proof.} We freeze the seed through the same map. The remaining density is
\begin{equation}
R(\mathcal J,\ell)=\frac{Q\ell^2+(P-Q)\ell}{\mathcal J}+\frac{P+Q}{2\ell^2},
\end{equation}
with $R(1,1)=(3P+Q)/2$. The domain is unchanged, so $\int\mathcal J d^3a=\int d^3a$, and we may add $P(\mathcal J-1)$. With $G=R+P\mathcal J-(5P+Q)/2$, $q=Q/P\in[0,1)$ and $A_q=q\ell^2+(1-q)\ell$, the arithmetic--geometric mean inequality and $\sqrt{A_q}\geq q\ell+(1-q)\sqrt\ell$ give
\begin{equation}
\frac GP\geq2\sqrt{A_q}+\frac{1+q}{2\ell^2}-\frac{5+q}{2}
\geq q\,\frac{(\ell-1)^2(2\ell+1)}{\ell^2}+(1-q)\frac{(x-1)^2(4x^3+3x^2+2x+1)}{2x^4}\geq0,
\end{equation}
with $x=\sqrt\ell$. Equality requires $\ell=\mathcal J=1$. For the seed, along an ideal path with $\mathbf v=0$ on the boundary, $\nabla\times\B_0=0$ gives $d\int\B_0\cdot\B_{0,t}dV/dt=0$, with no single-valued scalar potential needed. Hence
\begin{equation}
E_{\rm CGL}[\mathbf X]-E_{\rm CGL}[\mathrm{id}]
=\frac1{2\mu_0}\int|\B_{0,\mathbf X}-\B_0|^2dV+\int G\,d^3a\geq0.
\label{globalmin}
\end{equation}

\textbf{Lyapunov functional and Taylor relaxation.} Adding the kinetic energy $\frac12\int\rho_0|\dot{\mathbf X}|^2d^3a$ gives a nonnegative functional that smooth ideal CGL motion conserves. It bounds the kinetic energy and $\int|\B_{0,\mathbf X}-\B_0|^2dV$ while the motion stays smooth. Taylor relaxation (Ref.~\cite{taylorrelax}) fixes magnetic helicity and gives a force-free field. Here the material flux topology, the mass and the two CGL invariants are fixed instead, and the minimum carries perpendicular current and anisotropic pressure.

\section{Electrostatic stability}\label{app:8}
We fix the magnetic field and use the Hamiltonian guiding-center--Poisson model with $\mu$ conserved, an invariant phase-space measure $d\Gamma_s$ and a closed, conservative boundary that does no work. The equilibrium is $f_{0s}=F_s(\mathcal E_s,\mu)$ with $\phi_0=0$ and charge balance, and the energy is
\begin{equation}
\mathcal H_{\rm es}=\sum_s\int\mathcal E_s f_s\,d\Gamma_s+\frac{\epsilon_0}{2}\int|\nabla\phi|^2dV,\qquad
-\epsilon_0\nabla^2\phi=\sum_s q_s\int f_s\,d\Gamma_{v,s}.
\end{equation}

\textbf{Smooth decreasing distributions.} For $F_{s,\mathcal E}<0$, the unperturbed advection $L_s$ is antisymmetric with $L_sF_{s,\mathcal E}=0$, and $\partial_t\delta f_s+L_s\delta f_s=q_sF_{s,\mathcal E}L_s\delta\phi$. Multiplying by $\delta f_s/(-F_{s,\mathcal E})$ and integrating gives a rate $\int\delta\mathbf j\cdot\delta\mathbf E\,dV$, and the field energy changes at the opposite rate. Hence equation~\eqref{esenergy} is conserved and positive, and no mode with finite norm grows. The convex Casimirs $C_s$ with $C_{s,f}(F_s,\mu)=-\mathcal E$ and $C_{s,ff}=-1/F_{s,\mathcal E}$ give the conserved relative energy--Casimir functional
\begin{equation}
\mathcal L_{\rm es}=\sum_s\int\left[C_s(f_s)-C_s(f_{0s})-C_{s,f}(f_{0s})(f_s-f_{0s})\right]d\Gamma_s+\frac{\epsilon_0}{2}\int|\nabla\phi|^2dV\geq0,
\end{equation}
whose quadratic part is that invariant.

\textbf{Steps and plateaus.} The distributions of Appendix~\ref{app:4} need not be strictly decreasing. At each $\mu$ the phase-space volume of every superlevel set $\{f_s>a\}$ is conserved, and filling the lowest energies minimizes $\int\mathcal E_sd\Gamma_s$ at fixed volume. The superlevel sets of a non-increasing $F_s$ already have this form, so integration over $a$ gives $\sum_s\int\mathcal E_s(f_s-f_{0s})d\Gamma_s\geq0$ for every such rearrangement. With the field energy added,
\begin{equation}
\frac{\epsilon_0}{2}\int|\nabla\phi(t)|^2dV\leq\mathcal H_{\rm es}[f(0),\phi(0)]-\mathcal H_{\rm es}[f_0,0].
\end{equation}

\section{Action contours and rotational transform of the \texorpdfstring{$N=3$}{N=3} configuration}\label{app:9}
At fixed $(\mathcal E,\mu)$ the bounce-averaged drift over the labels $(q_1,q_2)$ of nearby trapped wells is $\dot q_1=C\partial_{q_2}J$, $\dot q_2=-C\partial_{q_1}J$, with a smooth nonzero $C$, so $\dot J=0$. A definite nondegenerate Hessian at a stationary point gives nearby closed contours, as long as the mirror points stay simple and the well persists.

At $B_m=29.9$, $30$ and $30.1$ the reduced action $j$ has minima $0.1488355$, $0.1545035$ and $0.1601559$. The smallest Hessian eigenvalues are $0.760$, $0.762$ and $0.764$, and the largest are $52.0$, $52.4$ and $52.9$. The sixteen contour roots lie on surface labels $[20.3555,20.3709]$, $[20.3654,20.3809]$ and $[20.3753,20.3908]$, with refinement residuals below $1.2\times10^{-9}$. Over all 32 mirror strengths $B_m=28.5,28.6,\dots,31.6$ of Fig.~\ref{fig:action}b, the smallest eigenvalue rises from $0.737$ to $0.798$ and the largest from $48.0$ to $63.2$. Every sampled boundary gap is positive, the refinement residuals stay below $3.6\times10^{-9}$, and the contour labels move outward from $[20.213,20.228]$ to $[20.515,20.530]$. The magnetic axis has label 19.

Field lines of the exact seed, traced over 160 periods and averaged with a weighted Birkhoff average, give rotational transforms $0.1968195$, $0.1915942$, $0.1846528$ and $0.1837743$ on the labels $19.6$, $20.0$, $20.355$ and $20.39$. The axis value is $\sqrt2/7=0.2020305$. Half-length averages agree to $4\times10^{-6}$ on the labels $20.355$ and $20.39$. The transform therefore falls by $8.6$ to $9.0\%$ from the axis to the surfaces of the contours in Fig.~\ref{fig:action}a, and this configuration is sheared.

\end{document}